\documentclass[a4paper,11pt]{article}
\usepackage{graphicx}
\usepackage{multicol}
\usepackage{subfigure}
\usepackage{graphicx,epsfig}
\usepackage{epstopdf}
\usepackage{latexsym}
\usepackage{amsfonts,amsmath,amssymb}
\usepackage{hyperref}
\usepackage{float}

\usepackage{epsfig,multicol,bbm}

\newcommand\fverb{\setbox\pippobox=\hbox\bgroup\verb}
\newcommand\fverbit{\egroup\item[\fbox{\unhbox\pippobox}]}

\newbox\pippobox

\begin{document}
\title{\bf Vacuum  $ f(R)$ thick brane solution with a Gaussian warp function }
\author{S. Sedigheh Hashemi   \,\,and \,\,  Nematollah Riazi\thanks{Electronic address: n\_riazi@sbu.ac.ir} 
\\
\small Department of Physics, Shahid Beheshti University, G.C., Evin, Tehran 19839,  Iran}
\maketitle
\begin{abstract}
This work deals with  $f(R)$  modified gravity in five dimensional space-time. The Gaussian thick brane is shown to be an exact solution in the frame work of $f(R)$ gravity in five dimensions with a bulk cosmological constant. Response of the brane to gravitational fluctuations and concordance with the Starobinsky  model is addressed. It is shown that the matter which supports the  Starobinsky $f(R)$ solution with the background geometry being flat FLRW with a Gaussian warp function, behaves like a radiation dominated era of universe, gradually changing to a dark energy dominated era.
\end{abstract}

\section{Introduction}\label{1}
In the past two decades, the brane world concept has been the subject of intensive investigations  in connection with the recent developments in superstring/M-theories\cite{horava}. The investigations were first  initiated in the work by Kaluza and Klein in $1920$s in order to unify two fundamental forces  electromagnetism and gravitation within the framework of a unified five-dimensional  theory. In this model, the size of the extra dimensions are compacted to the Planck scale\cite{plank}. However, in the brane world model, the sizes of the extra dimensions are about a few ${\rm Tev}^{-1}$\cite{3}-\cite{5}, millimeters \cite{6} or very large \cite{7,8}. 

According to the brane world  model, the standard-model particles are confined to a hypersurface, called a brane, immersed in a higher-dimensional spacetime called the bulk.
It is postulated that the matter fields are in the brane while the gravitational waves are free to propagate into the bulk.
The success of extra dimensions has  brought a solution for a number of insoluble problems in high-energy physics: the problem of mass hierarchy, stability, etc. This idea is carried by many theories, but the main  ones in this context are the one proposed by Arkani-Hamed, Dimopoulos and Dvali (ADD) \cite{6}, \cite{10, 11} and the so called, Randall-Sundrum (RS) model \cite{12, 13}. Particularly, the RS model has been advocated as a simple one.

In fact there are two RS models within the same framework.
In the RS-I model, the extra dimension appears due to the anti-de-Sitter (AdS) geometry along the fifth dimension. On the other hand, this model deals with two $D3$ branes on the $S_1/Z_2$ orbifold along the extra dimension \cite{ahmad}. The presence of two singular $D3$ branes with opposite tensions is needed for this model. It should be emphasized that, with this model the hierarchy problem can be solved without restoring to  the large campactified volume of the extra-dimension as proposed by ADD.

 In the RS-II model, the authors considered a $3$-brane (the four-dimensional Minkowski spacetime)  with a positive tension embedded in a five-dimensional anti de Sitter ($AdS_{5}$) spacetime. They showed that there exists a massless graviton (zero mode) and massive gravitons (Kaluza-Klein modes). The Newtonian gravity on the $3$-brane  is reproduced by the massless graviton. Therefore, by intuition, the massive modes which are the effect of the existence of the extra dimensions, cause a  correction to the Newtonian gravity. They also showed that in the low energy limit the Newtonian gravity can be recovered \cite{14}, \cite{15}. Since there is only one $D3$ brane, this model can not address the hierarchy problem\cite{ahmad}.

The brane should have some thickness which yields new possibilities and new problems\cite{16}. This kind of brane should fulfill  two main requirements. One is that the solutions should be regular and asymptotically flat, or de Sitter, the other is that the matter should be  restricted close to the brane. In the RS-II model, the 3-brane has no thickness, and the geometry has a singularity at the brane location. In order to escape the singularity, the extension of RS-II  model by replacing the 3-brane by a smooth thick brane obtained from a background scalar field can be invoked\cite{17}-\cite{19}. With this configuration of the thick brane, the bulk is not an $AdS_{5}$ spacetime\cite{20}.
Another way for generating thick branes instead of using scalar fields is to build them from pure geometry \cite{21}-\cite{24}.
  In these papers,  the gravitational zero mode as     well as the decoupling of the massive Kaluza-Klein modes are investigated.

In this paper, we use pure geometry for generating thick branes by invoking $f(R)$ theory where the gravitational Lagrangian is a function of Ricci scalar. The $f(R)$ theory was first created for studying the evolution  of the universe\cite{25}-\cite{27}. In the works \cite{28, 29}, the authors consider thick RS-II brane world solutions in pure $f(R)$ gravity. In \cite{28}, numerical solutions  obtained. Also an analytical thick brane solution is given in \cite{29}.

In this work, we derive a pure $f(R)$ solution by supposing a Gaussian thick brane.
The model  will be presented in the next section. In section \ref{3}, the gravitational fluctuations and the localization of gravity in the vicinity of the brane are discussed. In section \ref{4}, the $f(R)$ gravity in the Einstein frame is investigated and the corresponding scale factor and scalar potential versus the scalar curvature are  derived. In section \ref{5}, we consider the Starobinsky $f(R)$ model with the background geometry being flat FLRW universe with a Gaussian warp function and we study the behavior of the matter which supports the solution.

\section{The Model and the $f(R) $ Solution}\label{2}
We begin with  considering a pure $f(R)$ Lagrangian  which is an analytic function of Ricci scalar  in a five dimensional spacetime. The action specifying the dynamics of the brane-bulk system without matter source is\cite{fr}
\begin{equation}\label{1}
S=\frac{1}{2\kappa_5^2}\int {\rm d}^4x{\rm d}y \sqrt{-g}f(R),
\end{equation}
we use $\kappa_5^2=\frac{8\pi}{M_*^3}$, where $M_*$ is the five dimensional  Planck scale, $y$ denotes the extra dimension and $g$ is the determinant of the five dimensional   metric. We consider the flat and static brane embedded in a five dimensional  bulk which has the following metric
\begin{equation}\label{2}
{\rm d}s^2=e^{2A(y)}\eta_{\alpha \beta}{\rm d}x^\alpha{\rm d}x^\beta+{\rm d}y^2,
\end{equation}
where $e^{2A(y)}$ is the warp function and $\eta_{\alpha \beta}$ is the four dimensional  Minkowski metric with signature $(-,+,+,+)$. Throughout  this paper, Greek letters $\alpha, \beta$ run over $0, 1, 2, 3$ and capital Latin ones $A, B=0,1, 2, 3, 4$ are used to represent the brane and bulk indices, respectively.
 In the present case, we choose the warp function  $e^{2A(y)}= e^{-\lambda y^2}$ which has a Gaussian shape with $Z_2$ symmetry and $\sqrt{\lambda}$ is the inverse of the brane thickness $\Delta=\frac{1}{\sqrt{\lambda}}$.

To obtain the equations of motion, one can vary the action (\ref{1}) in the  usual manner  which gives 
\begin{eqnarray}
  R_{AB}F(R)-\frac12g_{AB}f(R)+(g_{AB}\square^{(5)}-\nabla_A\nabla_B)F(R)=0,
\label{eqEE}
\end{eqnarray} 
where  $\square^{(5)}=g^{AB}\nabla_{A}\nabla_{B}$ is the five-dimensional d'Alembert operator and $F(R)\equiv \frac{{\rm d}f(R)}{{\rm d}R}$.
By inserting metric (\ref{2}) in (\ref{eqEE}),  the following  field equations in the absence of matter can be obtained
\begin{equation}\label{3}
f(R)+2\lambda(4\lambda y^2-1)F(R)+6\lambda y \dot{F}(R)-2\ddot{F}(R)=0,
\end{equation}
and
\begin{equation}\label{4}
-8\lambda(\lambda y^2-1)F\left(R\right)-8\lambda y \dot{F}(R)-f(R)=0,
\end{equation}
where  dot stands for the derivative with respect to $y$. Adding  the above equations one can obtain 
\begin{equation}\label{5}
\ddot{F}+\lambda y \dot{F}-3 \lambda F=0,
\end{equation}
which is a second order differential equation for $F(R(y))$. Note that the Ricci scalar for metric (\ref{2}) and  the mentioned Gaussian warp function  is given by
\begin{equation}
R=-4\lambda (5\lambda y^2-2),
\end{equation}
which gives 
\begin{equation}
y=\pm \frac{1}{10\lambda}\sqrt{40\lambda -5R}
\end{equation}
 Hence, by solving Eq. (\ref{5}), the function $F(R(y))$ can be explicitly obtained given by
\begin{equation}\label{f}
F(R(y))=c_1 y (3+\lambda y^2)+c_2 e^{-\lambda y^2/2} \text{hypergeom}\left([2], [\frac{1}{2}], \frac{1}{2}\lambda y^2\right).
\end{equation}
Substituting $F(R(y))$ into Eq. (\ref{4}), $f(R(y))$  and consequently $f(R)$ can be calculated and is given by
\begin{align}
f(R)= & C(8\lambda -R)^{3/2}\left[1+\frac{1}{100\lambda}(8\lambda -R)\right]+8c_2\lambda e^{\frac{R-8\lambda}{40 \lambda}}\text{hypergeom}\left([2], [\frac{1}{2}], \frac{8\lambda-R}{40\lambda} \right)\nonumber\\&-\frac{8}{5}c_2(8\lambda -R) e^{\frac{R-8\lambda}{40 \lambda}}\text{hypergeom}\left([3], [\frac{3}{2}], \frac{8\lambda-R}{40\lambda} \right),
\end{align}
in which  $C\equiv \pm c_1 \frac{\sqrt{5}}{5\lambda}$, and the \lq\lq{}$\pm$\rq\rq{} signs account for two possible branches of solutions ($y=\pm \frac{1}{10\lambda}\sqrt{40\lambda -5R}$). Here we shall assume  $c_2=0$,  which  leads to
\begin{equation}\label{11}
f(R)=C(8\lambda -R)^{3/2}\left[1+\frac{1}{100\lambda}(8\lambda -R)\right].
\end{equation}
The  expansion of  $f(R)$ around $R= 0 $ up to the third order is
\begin{equation}\label{12}
f(R)=\frac{432}{25}\sqrt{2}C\lambda^{\frac{3}{2}}-\frac{17}{5}\sqrt{2\lambda}CR+\frac{21 C}{80 \sqrt{2\lambda}}R^2,
\end{equation}
where for small curvature that is $R\rightarrow 0$, the $f(R)$ function goes to a constant value
\begin{equation}
\lim _{R\rightarrow 0}f(R)=\frac{432}{25}\sqrt{2}C\lambda^{\frac{3}{2}}.
\end{equation}
Note that Eq. (\ref{11}) sets a maximum curvature 
\begin{equation}
R_{max}=8\lambda.
\end{equation}
Inserting  Eq. (\ref{12}) into the action (\ref{1}) and comparing it         with the Einstein-Hilbert action with a cosmological constant $\Lambda $ \cite{Mart}, i.e.
\begin{equation}
S_{EH}=\frac{1}{2\kappa_5^2}\int {\rm d}^4x{\rm d}y \sqrt{-g}(R^{(5)}-2\Lambda_{5}),
 \end{equation}
  leads to the following constraints
\begin{equation}
\sqrt{2\lambda}C=-\frac{5}{17}, \quad \Lambda_5=\frac{216}{85}\lambda \quad \text{and}\quad  c_{1}=\pm \frac{25\sqrt{\lambda}}{17\sqrt{10}}.
\end{equation}
\section{Gravitational Fluctuations}\label{3}
In this section, we shall consider the gravitational fluctuations of the  metric (\ref{2}), following the usual formalism\cite{1010}.
\begin{equation}
{\rm d}s^2=e^{2A(y)}\left(\eta_{\alpha \beta}+h_{\alpha \beta} \right){\rm d}x^\alpha{\rm d}x^\beta+{\rm d}y^2,
\end{equation}
where $h_{\alpha \beta}=h_{\alpha \beta}(x^{\rho},y)$ is a fluctuation which depends on all coordinates. By defining $a(y)\equiv e^{A(y)}$, the following fluctuations for the Riemann tensor and the  Ricci scalar are obtained 
\begin{eqnarray}
  \delta R_{\alpha \beta}&=&-\frac{1}{2}(\square^{(4)} h_{\alpha \beta }
               +\partial _{\alpha }\partial _{\beta }h
               -\partial _{\beta }\partial _{\sigma }h_{\alpha }^{\sigma }
               -\partial _{\alpha }\partial _{\sigma }h_{\beta }^{\sigma })
               -2 a a' h_{\alpha \beta }'\nonumber\\
               &-&3 h_{\alpha \beta } a'^2
               -a h_{\alpha \beta } a''
               -\frac{a^2 h_{\alpha \beta}''}{2}
               -\frac{a \eta _{\alpha \beta } a' h'}{2},\nonumber\\
\delta R_{\alpha 5}&=&\frac12\partial_y(\partial_\lambda
h_\alpha^\lambda-\partial_\alpha h),\quad
  \delta R_{55}=-\frac{1}{2} \left(\frac{2 a' h'}{a}+h''\right),\nonumber\\
\delta R&=&\delta (g^{\alpha \beta}R_{\alpha \beta})=-\frac{\square^{(4)}
h}{a^2}+\frac{\partial _{\alpha}\partial _{\beta } h^{\alpha \beta }}{ a^2}
        -\frac{ a' }{a}5h'-h'',
        \label{10}
\end{eqnarray}
where $\square^{(4)}=\eta^{\alpha \beta}\partial_{\alpha}\partial_{\beta}$, is d'Alembert operator
in  four-dimensions, prime denotes derivative with respect to $y$ and  $h=\eta^{\alpha \beta}h_{\alpha
\beta}$ is the trace of the tensor perturbations.

\begin{figure*}
\begin{multicols}{2}
    \includegraphics[width=\linewidth]{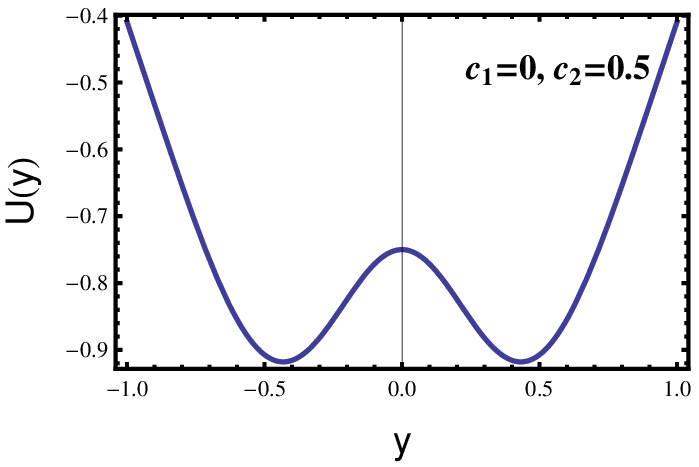}\par 
    \includegraphics[width=\linewidth]{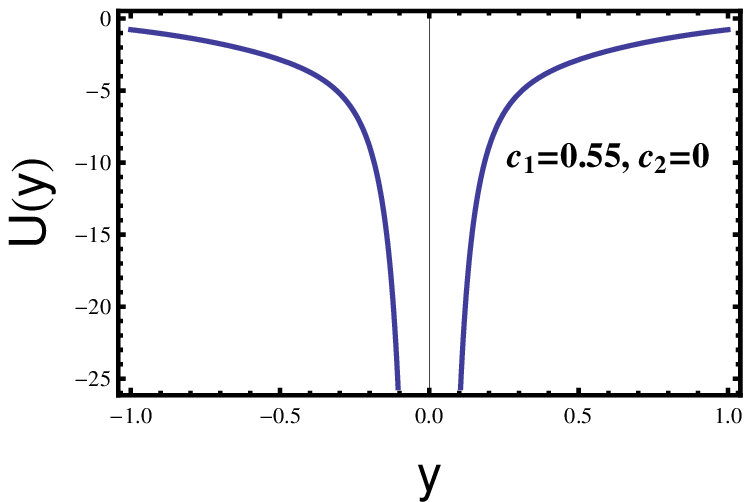}\par 
    \end{multicols}
\begin{multicols}{2}
    \includegraphics[width=\linewidth]{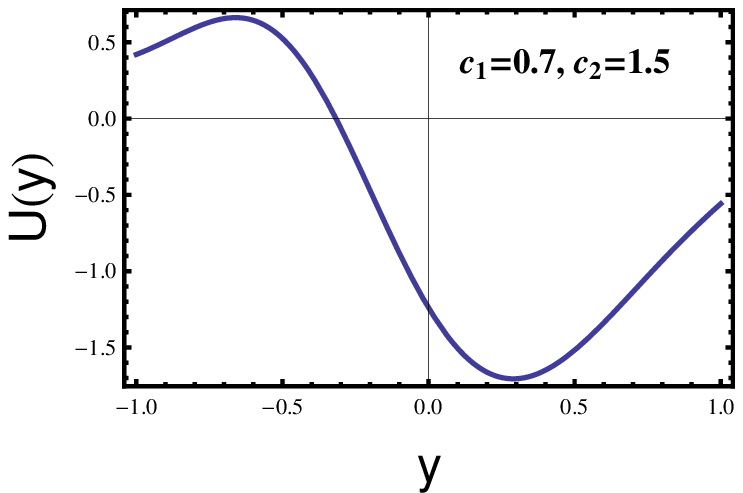}\par
    \includegraphics[width=\linewidth]{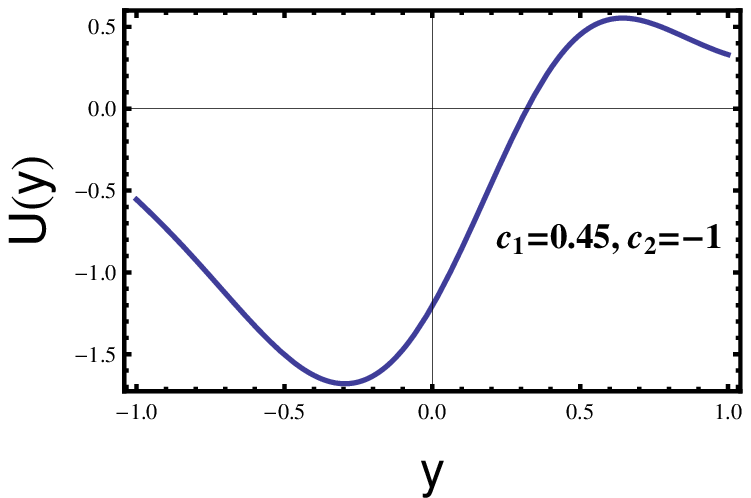}\par
\end{multicols}
\caption{\label{pot}The plots depict the behavior of the potential $U(y)$ with respect to the coordinate $y$. In all figures it is assumed that the brane thickness $\Delta=1$.}
\end{figure*}

In order to simplify the perturbed equations, we use the transverse-traceless gauge given by
\begin{eqnarray}
h=0=\partial_\mu h^\mu_{~\nu}\label{TTcondision}.
\end{eqnarray} 
With this choice, only $\delta
R_{\mu\nu}$ will not vanish.
The perturbation along the $f(R)$ equations of motion  (\ref{eqEE}) reads 
\begin{eqnarray}
 &&\delta R_{AB}F(R)+R_{AB}F(R)_{,R}\delta R-\frac12\delta g_{AB}f(R)
 -\frac12g_{AB}F(R)\delta R\nonumber\\
 &+&\delta(g_{AB}\square^{(5)}F(R))-\delta(\nabla_A\nabla_BF(R))=0.
 \label{eqPertubedEE}
\end{eqnarray}
For the above equations, we have 
\begin{eqnarray}
\nabla_A\nabla_BF(R)&=&(\partial_A\partial_B-\Gamma^P_{AB}\partial_P)F(R),\nonumber\\
g_{AB}\square^{(5)} F(R)&=&g_{AB}g^{MN}(\nabla_M\nabla_N F(R)),
\end{eqnarray}
and also 
\begin{eqnarray}
  \delta(\nabla_A\nabla_BF(R))&=&(\partial_A\partial_B-\Gamma^P_{AB}\partial_P)(F(R)_{,R}\delta
  R)-\delta \Gamma^P_{AB}\partial_P F(R),\\
  \delta(g_{AB}\square^{(5)} F(R))&=&\delta g_{AB}\square^{(5)} F(R)
  +g_{AB} \delta g^{MN}(\nabla_M\nabla_N F(R))\nonumber\\
  &+&g_{AB}g^{MN}\delta(\nabla_M\nabla_NF(R)).\label{34}
\end{eqnarray}
By using the transverse and traceless gauge, which leads to  $\delta R=0$, 
the above equations will become
\begin{eqnarray}
  \delta(\nabla_A\nabla_BF(R))&=&-\delta^\rho_A\delta^\sigma_B\delta\Gamma^5_{\rho\sigma} F\rq{}(R)~,\nonumber\\
  \delta(g_{AB}\square^{(5)} F(R))&=&\delta^\rho_A\delta^\sigma_B\delta g_{\rho\sigma} \square^{(5)}
F(R).
\end{eqnarray}
 Therefore, in  this gauge, we have
\begin{eqnarray}
  &&\delta(g_{AB}\square^{(5)} F(R))-\delta(\nabla_A\nabla_BF(R))\nonumber\\
&=&\delta^\rho_A\delta^\sigma_Ba^{2}\left[h_{\rho\sigma}\left(3\frac{a'}{a}F'(R)+F''(R)\right)
   -\frac12 F\rq{}(R)h_{\rho\sigma}'\right].
   \label{20}
\end{eqnarray}
The perturbed $f(R)$ equations of motion  (\ref{eqPertubedEE}) reduce to
\begin{eqnarray}\label{R}
  &&\delta R_{AB}F(R)-\frac12\delta g_{AB}f(R)\nonumber\\
 &+&\delta^\rho_A\delta^\sigma_B a^{2}\left[h_{\rho\sigma}\left(3\frac{a'}{a}F\rq{}(R)+F''(R)\right)
   -\frac12h_{\rho\sigma}'F\rq{}(R)\right]=0.
   \label{21}
\end{eqnarray}
By plugging (\ref{10})  into (\ref{R}), we obtain the 
$(\alpha,\beta)$ components of the perturbed   $f(R)$ equations as 
\begin{eqnarray}
  &&\left(-\frac{1}{2}\square^{(4)} h_{\alpha \beta}
               -3 h_{\alpha \beta }a'^2
               -2 a a' h_{\alpha \beta }'
               -a h_{\alpha \beta } a''
               -\frac{a^2 h_{\alpha \beta }''}{2 }\right)F(R)\nonumber\\
               &&-\frac12a^2h_{\alpha \beta }f(R)
               +a^{2}\left[h_{\alpha \beta }\left(3\frac{a'}{a}F\rq{}(R)+F''(R)\right)
   -\frac12h_{\alpha \beta }'F\rq{}(R)\right]=0.
  \label{A38}
\end{eqnarray}
On the other hand, the $(\alpha,~\alpha)$ components of $f(R)$ equations
(\ref{eqEE}) is
\begin{eqnarray}
  f(R)+2F(R)\left[3\left(\frac{a'}{a}\right)^2+\frac{a''}{a}\right]
  -6F\rq{}(R)\frac{a'}{a}-2F''(R)=0.
  \label{A39}
\end{eqnarray} 
By   simplifying  Eq.~(\ref{A38}), one can get 
\begin{eqnarray}
  &&\left(-\frac{1}{2}\square^{(4)} h_{\alpha \beta  }
               -2 a a' h_{\alpha \beta  }'
               -\frac{a^2 h_{\alpha \beta  }''}{2 }\right)F(R)
               -\frac12a^{2}h_{\alpha \beta }'F(R)=0.
\end{eqnarray}
Consequently, the $(\alpha,~\beta)$
components of the perturbed $f(R)$ equations  read
\begin{eqnarray}
  &&\left(a^{-2}\square^{(4)} h_{\alpha \beta }
               +4 \frac {a'}{a} h_{\alpha \beta }'
               + h_{\alpha \beta }''
               \right)F(R)+h_{\alpha \beta}'F\rq{}(R)=0.
\end{eqnarray}
This can be written as
\begin{eqnarray}
\square^{(5)}h_{\alpha \beta}=\frac{F\rq{}(R)}{F(R)}\partial_y h_{\alpha \beta}.
\label{eqEEFinal}
\end{eqnarray}
Introducing a coordinate transformation
\begin{eqnarray}
dz=a^{-1}dy,\label{coordinatetrans}
\end{eqnarray}
 the perturbed equation (\ref{eqEEFinal}) can be written as 
\begin{eqnarray}
 \left[\partial_z^{~2}
  +\left(3\frac{\partial_z a}{a}
  +\frac{\partial_z F(R)}{F(R)}\right)\partial_z
  +\square^{(4)}\right]h_{\alpha \beta}=0.
\end{eqnarray}
Following \cite{1010}, we look for solutions of the form 
$h_{\alpha \beta}(x^{\rho},z)=(a^{-3/2}F(R)^{-1/2})\epsilon_{\alpha \beta}(x^{\rho})\psi(z)$,
in which $\epsilon_{\alpha \beta}(x^{\rho})$ satisfies the transverse and
traceless condition $\eta^{\alpha \beta}\epsilon_{\alpha \beta}=0$ and $\partial_\alpha
\epsilon^{~\alpha}_\beta=0$. Then the we end up with   a Schr\"odinger  like equation for
$\psi(z)$ 
\begin{eqnarray}
  \left[\partial_z^2
      -U(z)\right]\psi(z)
      =-m^2\psi(z),\label{Schrodinger}
\end{eqnarray}
 where the potential $U(z)$ is given by
\begin{eqnarray}
 U(z)=\frac34\frac{a'^2}{a^2}
      +\frac32\frac{a''}{a}
      +\frac32\frac{a' F\rq{}(R)}{a F(R)}
      -\frac14\frac{F\rq{}(R)^2}{F(R)^2}
      +\frac12\frac{F''(R)}{F(R)}.
      \label{Schrodingerpotential}
\end{eqnarray}
 In order to understand the behavior of the potential $U(z)$, we  use the coordinate transformation (\ref{coordinatetrans}), and then obtain the potential as a function of the coordinate $y$. In Figure \ref{pot}, we show the potential $U(y)$ for different values of $c_1$ and $c_2$ which are used in equation (\ref{f}).
The minimum of the potential is related to the stability of the solution. For the case with $c_1=0$ and $c_2=0.5$, there are two stable points. With $c_1=0.55$ and $c_2=0$, the potential is singular at  $y=0$. For $c_1=0.7, c_2=1.5$ and $c_1=0.45, c_2=-1$, the potential has only one stable point.


One can also factorize the Schr\"odinger like 
equation (\ref{Schrodinger}) as
\begin{eqnarray}
 \left[\left(\partial _z
 +\left(\frac{3}{2}\frac{\partial_z a}{a}+\frac{1}{2}\frac{\partial_z F(R)}{F(R)}\right)\right)
 \left(\partial_z
 -\left(\frac{3}{2}\frac{\partial_z a}{a}+\frac{1}{2}\frac{\partial_z F(R)}{F(R)}\right)\right)\right]\psi(z)
 =-m^2\psi(z),
\end{eqnarray} 
which shows  that there
is no gravitational mode with $m^2<0$. As a result  any solution of Eqs. 
 (\ref{1}), (\ref{2}) is stable under the tensor
perturbations. If the zero mode exists, it will have the form 
\begin{eqnarray}
\psi^{(0)}(z)=N_0 a^{3/2}(z)\sqrt{F(z)}, \label{zeromode}
\end{eqnarray} with $N_0$ the normalization constant. A normalizable $\psi^{(0)}(z)$  leads to the Newton\rq{}s law in four-dimensional gravity \cite{8}, \cite{G}.
The zero mode $\psi^{(0)}(z)$  is normalizable if
\begin{equation}
1= \int ^{+\infty}_{-\infty}|\psi^{(0)}(z)|^2{\rm d}z=N_0^2\int ^{+\infty}_{-\infty}e^{3A(z)}F(R(z)){\rm d}z=N_0^2\int ^{+\infty}_{-\infty}e^{2A(y)}F(R(y)){\rm d}y,
\end{equation}
can be satisfied, which for our case it can not  be integrated analytically. However, the numerical integration with the use of Eq. (\ref{f}) gives the value $N_0=0.5541$,  which confirms that for our  solution, the gravitational zero mode is normalizable and can be localized on the brane. In this case, the Newton\rq{}s law can be retrieved on the brane.

In  general relativity,
Eq.~(\ref{eqEEFinal}) will  reduce to the five-dimensional Klein-Gordon
equation for the massless spin-2 gravitons. Nevertheless, by having an
arbitrary function  $f(R)$ and non-constant curvature $R$, the
equation for $h_{\alpha \beta}$ is completely different from the massless
Klein-Gorden equation.
Moreover,  by using the  transverse and traceless gauge, the perturbed equation always remains second order.
 Some of  the application of these  results
 are  given  in~\cite{Afonso2007}.

\section{ Einstein frame}\label{4}
In this section, in order to improve our understanding about the   dynamics of the  $f(R)$ model  we transform to the Einstein frame. Practically, it is not easy to define the scalar degree of freedom for a $f(R)$ gravity. Instead, using the effective potential in a particular form will be convenient \cite{45}. We use the following conformal transformation for switching to the Einstein frame \cite{27}, \cite{35}, \cite{38}
\begin{equation}
\tilde{g}_{AB}=\Omega^2 g_{AB},
\end{equation}
where $\Omega^2$ is the conformal factor and a tilde represents a quantity in Einstein frame. Under the conformal translation, the Ricci scalar transforms as \cite{38}
\begin{equation}\label{41}
R=\Omega^2\tilde{R}+8\tilde{g}^{AB}\Omega(\tilde{\nabla}_{A}\tilde{\nabla}_{B}\Omega)-20\tilde{g}^{AB}\tilde{\nabla_{A}}\Omega\tilde{\nabla}_{N}\Omega,
\end{equation}
where $\tilde{\nabla}_{A}$ is the covariant derivative in the conformal frame.

We  can therefore rewrite the action (\ref{1}) in the form 
\begin{equation}\label{s}
S=\int {\rm d}^4x{\rm d}y\sqrt{-\tilde{g}}\left(\frac{F(R)}{2\kappa_5^2}\Omega^{-5}R-\Omega^{-5}U\right),
\end{equation}
where 
\begin{equation}
U=\frac{F(R)R-f(R)}{2\kappa_5^2}.
\end{equation}
In the above equations we  used $\sqrt{-g}=\Omega^{-5}\sqrt{-\tilde{g}}$. Substituting Eq. (\ref{41}) into  (\ref{s}) and  making  the  choice $F(R)=\Omega^3$, then the action will reduce to
\begin{equation}\label{s}
S=\int {\rm d}^4x{\rm d}y\sqrt{-\tilde{g}}\left(\frac{1}{2\kappa_5^2}[\tilde{R}-12\Omega^{-2}\tilde{g}^{AB}\tilde{\nabla}_{A}\Omega\tilde{\nabla}_{B}\Omega]-\Omega^{-5}U\right).
\end{equation}
By defining 
\begin{equation}\label{45}
\phi=2\sqrt{\frac{3}{\kappa_5^2}}\ln {\Omega}, \quad V(\phi)=\Omega^{-5} U,
\end{equation}
the action is  simplified  as
\begin{equation}
S=\int {\rm d}^4x{\rm d}y\sqrt{-\tilde{g}}\left(\frac{1}{2\kappa_5^2}\tilde{R}-\frac{1}{2}\tilde{g}^{AB}\tilde{\nabla}_{A}\phi\tilde{\nabla}_{B}\phi -V(\phi)\right).
\end{equation}
For our model the conformal factor is
\begin{equation}
\Omega^2=F(R)^{2/3}=\exp{(\frac{\kappa \phi}{\sqrt{3}})}.
\end{equation}

Using the Eq. (\ref{45}), the scalar field corresponding to our model (\ref{11}) can be expressed as
\begin{equation}\label{47}
\phi=\frac{2}{3}\sqrt{\frac{3	}{\kappa_5^2}}\ln\left({\frac{C(68\lambda-R)\sqrt{8\lambda-R}}{40\lambda}}\right).
\end{equation}
 Solving Eq. (\ref{47}) to obtain $R$ versus $\phi$ gives
\begin{align}
R=2\left(24\lambda+ 10^{4/3}\frac{C^2\lambda^2}{\Phi(\phi)}+10^{2/3}\frac{\Phi(\phi)}{C^2}\right),
\end{align}
where $A$ is defined as $A \equiv \frac{2}{\sqrt{3}\kappa_5}$ and $\Phi(\phi)$  is given by
\begin{equation}
\Phi(\phi)\equiv    \left[-C^4\lambda^2 e^{2\phi/A}-10C^6\lambda^3+\sqrt{\lambda^4 C^8 e^{2\phi/A}      (e^{2\phi/A}  +20\lambda C^2 ) }\right   ]^{1/3}.
\end{equation}
Using this expression for $R$, the $f(R(\phi))$ and consequently the field potential (\ref{45}) takes the form
\begin{align}
V(\phi)= &\frac{e^{-2\phi/3A}}{\kappa_{5}^2} \left(       24\lambda  +10^{4/3} \frac{ \Phi(\phi)}{C^2} +10^{2/3}\frac{C^2\lambda^2}{\Phi(\phi)}\right) 
\nonumber\\&-\frac{e^{-5\phi/3A}}{\kappa_{5}^2} \sqrt{2}C\left(      -20\lambda- 10^{4/3} \frac{ \Phi(\phi)}{C^2} -10^{2/3}\frac{C^2\lambda^2}{\Phi(\phi)}                                                           \right)^{3/2}\times \nonumber\\& \left(1-\frac{20C^2\lambda+\frac{10^{4/3}C^4\lambda^2}{\Phi(\phi)}       +10^{2/3}\Phi     (\phi)  }{50C^2}\right).
\end{align}

\section{Flat FLRW Brane with  Starobinsky $f(R)$ model}\label{5}
In this section, we consider a flat FLRW brane with a scale factor $a(t)$ depending on the cosmological time $t$ 
\begin{equation}\label{24}
{\rm d}s^2=e^{-\lambda y^2}\left[-{\rm d}t^2+a^2(t)({\rm d}r^2+r^2{\rm d}\Omega^2)\right]+{\rm d}y^2,
\end{equation}
the Ricci scalar for metric (\ref{24}) is given by
\begin{equation}
R=\frac{2}{a^2 e^{-\lambda y^2}}\left[   3a\ddot{a}+3\dot{a}^2-10a^2\lambda^2y^2e^{-\lambda y^2}+4a^2\lambda e^{-\lambda y^2}               \right],
\end{equation}
where dot represents the time derivative. As it is seen, the Ricci scalar is a function of $t$ and $y$. Here, we restrict ourselves to cases where the    Ricci scalar has only $y$ dependence. Hence, by putting the time dependent part of the Ricci scalar equals zero, that is $a(t)\ddot{a}(t)+\dot{a}(t)^2=0$, we will find $a(t)=\pm \sqrt{2c_1 t+2c_2}$ which has a radiation like behavior, $c_1$ and $c_2$ being  constants of integration. We put $c_2=0$ in order to have the same reference time for the big bang as in FLRW models. Note that, in our model as the scale factor goes to zero $a(t) \rightarrow 0$, the Ricci scalar does not diverge which  is completely different with the FLRW cosmological model, where in the latter  the scalar curvature will diverge.
 
 Now with the above choice, the  metric (\ref{24}) reduces to
\begin{equation}\label{13}
{\rm d}s^2=e^{-\lambda y^2}\left[-{\rm d}t^2+2c_1 t({\rm d}r^2+r^2{\rm d}\Omega^2)\right]+{\rm d}y^2.
\end{equation}
Similar to the calculations of  section (\ref{1}), we begin with the action (\ref{1}) plus a matter term $S_M$, therefore the total action for $f(R)$ gravity takes the form 
\begin{equation}
S=\frac{1}{2\kappa_5^2}\int {\rm d}^4x{\rm d}y \sqrt{-g}f(R)+S_M(g_{AB}).
\end{equation}
Variation with respect to the metric $g_{AB}$ gives the $f(R)$ modified gravity as
\begin{equation}\label{17}
F(R)R_{AB}-\frac{1}{2}f(R)g_{AB}-\nabla_{A}\nabla_{B}F(R)+g_{AB}\Box F(R)=\kappa_{5}^2T_{AB}
\end{equation}
where $\Box \equiv \nabla^C \nabla_{C} $ is the d\rq{}Alembert  operator and
\begin{equation}
T_{AB}=-\frac{2}{\sqrt{-g}}\frac{\delta S_M}{\delta g^{AB}}.
\end{equation}
 
 In order to clarify explicitly the behavior of the above field equations, we consider a    perfect fluid which is  characterized by 
 \begin{equation}
 T_{AB}=(\rho+P)u_A u_B+Pg_{AB},
 \end{equation}
where $\rho$ is the energy density, $P$ is the pressure and $u_{A}$ is the velocity vector. Moreover, we assume that the  $f(R)$ function is the famous Starobinsky  $f(R)$ model given by
\begin{equation}
f(R)=R+\alpha R^2,
\end{equation}
which conforms with the expansion of our model up to $R^2$ and with $\Lambda=0$ in the present case.
Obviously $F(R)=1+2\alpha R$ and the Ricci scalar $R=-4\lambda(5\lambda y^2-2)$. As a consequence, $f(R)$ and $F(R)$ will  be functions of $y$.

We set up the field equations (\ref{17}) which have the following components
\begin{equation}\label{21}
F(R)R_{tt}-\frac{1}{2}f(R)g_{tt}+g_{tt}\Box F= \kappa_{5}^2 T_{tt}=\kappa_{5}^2 \rho e^{-\lambda y^2}
\end{equation}
\begin{equation}\label{22}
F(R)R_{rr}-\frac{1}{2}f(R)g_{rr}+g_{rr}\Box F= \kappa_{5}^2 T_{rr}=\kappa_{5}^2 T_{\theta \theta}=\kappa_{5}^2 T_{\phi \phi}=\kappa_{5}^2P_r e^{-\lambda y^2}
\end{equation}
and
\begin{equation}\label{37}
F(R)R_{yy}-\frac{1}{2}f(R)g_{yy}-\nabla_y\nabla_yF(R)+\Box F= \kappa_{5}^2 T_{yy}=\kappa_{5}^2P_y.
\end{equation}
The above equations  reduce to
\begin{equation}\label{42}
4F(R)H^2A(t,y)+\frac{1}{2}f(R)-\Box F(R)= \kappa_{5}^2\rho(t,y),
\end{equation}
\begin{equation}\label{43}
4F(R)H^2B(t,y)-\frac{1}{2}f(R)+\Box F(R)= \kappa_{5}^2P(t,y),
\end{equation}
and
\begin{equation}\label{62}
-40\alpha \lambda^4 y^4-64\alpha \lambda^3 y^2+6\lambda^2 y^2+32\alpha \lambda^2=\kappa_{5}^2P_y(y),
\end{equation}
where $A(t,y)$ and $B(t,y)$ are given by
\begin{equation}
A(t,y)=\left[\frac{3}{4}e^{\lambda y^2} +\lambda(4\lambda y^2-1)t^2 \right],
\end{equation}
\begin{equation}
B(t,y)=\left[ \frac{1}{4}e^{\lambda y^2} -\lambda(4\lambda y^2-1)t^2      \right],
\end{equation}
and $H$ is the Hubble parameter defined as $H=\dot{a}(t)/a(t)=\frac{1}{2t} $.
Inserting $f(R)$ and $F(R)$ as a function of $y$ in Eqs. (\ref{42}), (\ref{43}) we obtain
\begin{align}\label{64}
\kappa_{5}^2 \rho(t,y)=&4H^2\left(1-8\alpha\lambda(5\lambda y^2-2)\right)\left(\frac{3}{4}e^{\lambda y^2}+\lambda(4\lambda y^2-1)t^2         \right) \nonumber\\&+ (5\lambda y^2-2)\left(-2\lambda+8\lambda^2 \alpha(5\lambda y^2-2)\right)
+80\alpha \lambda^2(1+4\lambda y^2),
\end{align}
\begin{align}
\kappa_{5}^2 P(t,y)=&4H^2\left(1-8\alpha\lambda(5\lambda y^2-2)\right)\left(\frac{1}{4}e^{\lambda y^2}-\lambda(4\lambda y^2-1)t^2         \right) \nonumber\\&+ (5\lambda y^2-2)\left(2\lambda-8\lambda^2 \alpha(5\lambda y^2-2)\right)
-80\alpha \lambda^2(1+4\lambda y^2).
\end{align}
 As it is seen from (\ref{62}), the pressure in the $y$ direction is only a function of $y$ and independent of the  cosmic time $t$.

An effective equation of state (EoS) parameter can be introduced as 
\begin{equation}
w(t,y)=\frac{P(t,y)}{\rho(y,t)},
\end{equation}
which can be regarded as the dynamical EoS parameter of the model.

Furthermore, the energy conservation law $\nabla_A T^{A 0}=0$ gives
\begin{equation}
\dot{\rho}+\frac{3}{2t}P+\frac{3}{2t}\rho=0,
\end{equation}
which conforms with the FLRW continuity equation
\begin{equation}
\dot{\rho}+3H(\rho+P)=0,
\end{equation}
during radiation-dominated era where $H(t)=\frac{1}{2t}$.

In particular, given that
\begin{equation}
\frac{\ddot{a}}{a}=\dot{H}+H^2,
\end{equation}
the modified Friedman equations become
\begin{equation}
\left(\frac{\dot{a}}{a}\right)^2=\frac{\kappa_5^2 \rho}{4F(R)A(t,y)}-\frac{f(R)}{4F(R)A(t,y)}+\frac{\Box F(R)}{4F(R)A(t,y)},
\end{equation}
\begin{equation}
\frac{\ddot{a}}{a}=-\frac{\kappa_5^2}{8F(R)A(t,y)}(\rho+3P)-\frac{\dot{A}(t,y)H}{2A(t,y)}+\frac{\Box F(R)}{4F(R)A(t,y)}-\frac{f(R)}{8F(R)A(t,y)}.
\end{equation}
\\
Here we consider the dynamical equations on the  brane which is located at $y=0$. In this case,  the EoS parameter  reduces to
\begin{equation}\label{73}
w(t, y=0)=\frac{4(1+16\alpha \lambda)H^2(\frac{1}{4}     +\lambda t^2       )-4\lambda (1+28 \alpha \lambda )}{4(1+16\alpha \lambda)H^2(\frac{3}{4}-\lambda t^2       )+4\lambda(1+28\alpha \lambda)}.
\end{equation}

The behavior of the EoS parameter (\ref{73}) as a function  of $t$   is depicted in  Figure \ref{w1}. 
We immediately see from the figure that, the EoS parameter starts from $w=1/3$ which is the radiation equation of state parameter  and then for large values of $t$ it converges to a constant value $w=-1$ which corresponds to dark energy equation of state parameter.
We also plotted the energy density (\ref{64}) as a function of time on the brane in Figure {\ref{roh}. It is shown that the energy density decreases as the time goes on similar to radiation dominate era where the radiation obeys the standard continuity equation $(\rho_r\propto  a(t)^{-4})$, but with a difference that in this model the energy density arrives to a constant value for large $t$  which corresponds to energy density of the dark energy.

\begin{center}
\begin{figure}[H] \hspace{4.cm}\includegraphics[width=8.cm]{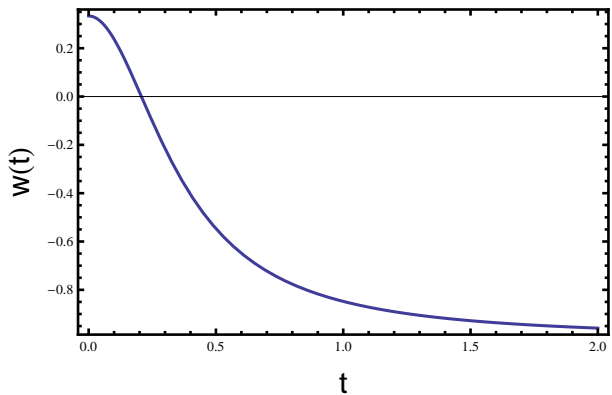}\caption{\label{w1} \small
The EoS parameter  $w$ as a function of $t$ on the brane($y=0$) for arbitrary values  $\Delta, \alpha=1$. It starts from $w=1/3$ to $w=-1$ which corresponds to radiation and dark energy EoS parameter, respectively.}
\end{figure}
\end{center}
\begin{center}
\begin{figure}[H] \hspace{4.cm}\includegraphics[width=8.cm]{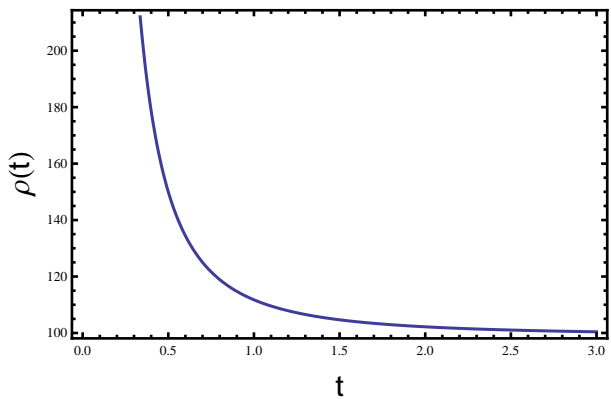}\caption{\label{roh} \small
Energy density as a function  of time for arbitrary values  $\Delta, \alpha=1$. It is seen that the brane energy density converges to a constant value.}
\end{figure}
\end{center}


\section{Conclusion}
In this paper, we have investigated braneworld models with a Gaussian warp function which has a $Z_2$ symmetry and also with a single extra spatial dimension of infinite extent. We replaced the Einstein-Hilbert Lagrangian by a nonlinear $f(R)$ Lagrangian in vacuum, including an extra dimension.  We worked out the $f(R)$ equations of motion, which lead to an  exact vacuum $f(R)$ solution. By appropriately setting constants of integration,
a bulk cosmological constant can be obtained.

We studied the gravitational fluctuations of our solution by adding small tensor perturbations. We realized that the solution is stable against the  perturbations. We also showed that the gravitational zero mode is normalizable and can be localized on the brane. The behavior of the potential was thoroughly addressed, with the  minimum of the potential  regarded as the stable point which leads to the desired stability.

Finally, we considered the flat FLRW brane model with a Gaussian warp factor and the $R^2$ approximation. We showed  that, the matter which supports the solution starts  like a radiation dominated era and  in the late time it acts like  dark energy with a constant energy density. The equation of state parameter was shown to  start from $w= 1/3$ (radiation)  and end with $w=-1$ (dark energy).


\begin{thebibliography}{99}
\bibitem{horava}
P. Horava and E. Witten, Nucl. Phys. B 460, 506
(1996); B475, 94 (1996);
E. Witten, Nucl. Phys. B 471, 135 (1996).
\bibitem{plank}
T. Appelquist, A. Chodos, P. G. O. Freund, Modern Kaluza-Klein Theories,
Addison-Wesley Publishing Company, (1987).
\bibitem{3}
I. Antoniadis, Phys. Lett. B 246, 377 (1990) .
\bibitem{4}
I. Antoniadis, A. Arvanitaki, S. Dimopoulos, A. Giveon, Phys. Rev. Lett. 108, 081602 (2012).
\bibitem{5}
K. Yang, Y.-X. Liu, Y. Zhong, X.-L. Du, S.-W. Wei, Phys. Rev. D 8, 127502 (2012).
\bibitem{6}
I. Antoniadis, N. Arkani-Hamed, S. Dimopoulos, G. Dvali, Phys. Lett. B
436, 257–263 (1998).
\bibitem{7}
V. A. Rubakov, M. E. Shaposhnikov, Phys. Lett. B 125, 136 (1983).
\bibitem{8}
L. Randall, R. Sundrum, Phys. Rev. Lett. 83, 4690 (1999) .
\bibitem{10}
N. Arkani-Hamed, S. Dimopoulos and G. R. Dvali, Phys. Lett. B 429, 263 (1998).
\bibitem{11}
N. Arkani-Hamed, S. Dimopoulos and G. R. Dvali, Phys. Rev. D 59, 086004 (1999) [hepph/9807344].
\bibitem{12}
L. Randall and R. Sundrum, Phys. Rev. Lett. 83, 3370 (1999) [arXiv:hep-ph/9905221].
\bibitem{13}
L. Randall and R. Sundrum, Phys. Rev. Lett. 83, 4690 (1999) [arXiv:hep-th/9906064].
\bibitem{ahmad}
A. Ahmed and B. Grzadkowski. arXiv preprint arXiv:1210.6708. 2012 Oct 25.
\bibitem{14}
J. Garriga and T. Tanaka, Phys. Rev. Lett. 84, 2778 (2000).
\bibitem{15}
S. B. Giddings, E. Katz and L. Randall, JHEP 0003, 023 (2000).
\bibitem{16}
V. Dzhunushaliev, V. Folomeev and M. Minamitsuji, arXiv: 0904.1775 [gr-qc].
\bibitem{17}
O. DeWolfe, D. Z. Freedman, S. S. Gubser, A. Karch, Phys. Rev. D 62, 046008
(2000).
\bibitem{18}
M. Gremm, Phys. Lett. B 478, 434 (2000).
\bibitem{19}
C. Csaki, J. Erlich, T. J. Hollowood, Y. Shirman, Nucl. Phys. B 581, 309–338 (2000).
\bibitem{20}
Y. Zhong and Y.-X. Liu,
Eur. Phys. J. C 76, 321 (2016), [arXiv: 1507.00630].
\bibitem{21}
O. Arias, R. Cardenas, I. Quiros, Nucl. Phys. B 643, 187 (2002).
\bibitem{22}
N. Barbosa-Cendejas, A. Herrera-Aguilar, J. High Energy Phys. 10, 101 (2005).
\bibitem{23}
N. Barbosa-Cendejas, A. Herrera-Aguilar, Phys. Rev. D 73, 084022 (2006).
\bibitem{24}
N. Barbosa-Cendejas, A. Herrera-Aguilar, M. A. Reyes Santos, C. Schubert,
Phys. Rev. D 77, 126013 (2008).
\bibitem{25}
 H. A. Buchdahl, Mon. Not. Roy. Astron. Soc. 150, 1 (1970).
\bibitem{26}
J. D. Barrow, A. C. Ottewill, J. Phys. A 16, 2757 (1983).
\bibitem{27}
J. D. Barrow, S. Cotsakis, Phys. Lett. B 214, 515 (1988).
\bibitem{28}
V. Dzhunushaliev, V. Folomeev, B. Kleihaus, J. Kunz, J. High Energy
Phys. 04, 130 (2010).
\bibitem{29}
H. Liu, H. Lu, Z.-L. Wang, J. High Energy Phys. 1202, 083 (2012).
\bibitem{fr}
De Felice, Antonio, and Shinji Tsujikawa.  Living Rev. Rel 13.3, 1002-4928 (2010).
\bibitem{Mart}
Maartens R. Brane-world gravity. arXiv preprint gr-qc/0312059. 2003 Dec 10.
\bibitem{1010}
Zhong, Y., Liu, Y. X., and Yang, K. Physics Letters B, 699(5), 398-402, (2011).
\bibitem{G}
M. Giovannini, Phys. Rev. D 64, 064023, (2001).
\bibitem{Afonso2007}
V.~I. Afonso, D.~Bazeia, R.~Menezes, A.~Y. Petrov
 Phys. Lett.B658, 71-76, (2007).
 \bibitem{45}
A. V. Frolov, Phys. Rev. Lett. 101, 061103 (2008).
\bibitem{35}
A. De Felice and S. Tsujikawa, Living Rev. Rel. 13, 3 (2010).
\bibitem{37}
K. Maeda, Phys. Rev. D 39, 3159 (1989).
\bibitem{38}
S. Carroll, Spacetime And Geometry An Introduction To General Relativity, Pearson Education, Inc., 2004.
\bibitem{star}
A. A. Starobinsky, Physics  Letters B, 91, 99-102 (1980).
 
 
 
 
 
 
 
 
 
 
 
 


\end{thebibliography}
\end{document}